\RequirePackage{fix-cm}
\documentclass[twocolumn]{svjour3}          
\smartqed  
\usepackage{graphicx}
\usepackage{amssymb,amsmath}
\usepackage{enumerate}
\usepackage[para,online,flushleft]{threeparttable}
\usepackage{multirow}
\usepackage{subfig}
\usepackage[linesnumberedhidden,ruled,vlined]{algorithm2e}
\usepackage{algpseudocode}
\usepackage{url}
\usepackage{hyperref}
\usepackage{color}

\let\oldnl\nl
\newcommand{\nonl}{\renewcommand{\nl}{\let\nl\oldnl}}

\newcommand{\mbR}{\mathbb{R}}
\begin{document}

\title{Low-Cost and High-Throughput Testing of COVID-19 Viruses and Antibodies via Compressed Sensing: System Concepts and Computational Experiments
}


\author{Jirong Yi\and
         Raghu Mudumbai \and
         Weiyu~Xu
}


\institute{ \small{Jirong Yi, Raghu Mudumbai and Weiyu Xu. \at
              Dept. of ECE, University of Iowa, Iowa City, IA, 52242 \\
              \email{weiyu-xu@uiowa.edu}}
}

\date{Received: date / Accepted: date}

\maketitle

\begin{abstract}
Coronavirus disease 2019 (COVID-19) is an ongoing pandemic infectious disease outbreak that has significantly harmed and threatened the health and lives of millions or even billions of people. COVID-19 has also negatively impacted the social and economic activities of many countries significantly. With no approved vaccine available at this moment, extensive testing of COVID-19 viruses in people are essential for disease diagnosis, virus spread confinement, contact tracing, and determining right conditions for people to return to normal economic activities.  Identifying  people who have antibodies for COVID-19 can also help select persons who are suitable for undertaking certain essential activities or returning to workforce. However, the throughputs of current testing technologies for COVID-19 viruses and antibodies are often quite limited, which are not sufficient for dealing with COVID-19 viruses' anticipated fast oscillating waves of spread affecting a significant portion of the earth's population.

In this paper, we propose to use compressed sensing (group testing can be seen as a special case of compressed sensing when it is applied to COVID-19 detection) to achieve high-throughput rapid testing of COVID-19 viruses and antibodies, which can potentially provide tens or even more folds of speedup compared with current testing technologies. The proposed compressed sensing system for high-throughput testing can utilize expander graph based compressed sensing matrices developed by us \cite{Weiyuexpander2007}.

\end{abstract}

\section{Introduction}

The ongoing Covid-19 pandemic has already claimed thousands of human lives. In addition, it has also forced a worldwide shutdown of social life and commerce, and the resulting economic depression has caused tremendous suffering for millions of people. 

In the absence of a vaccine, the experience of public health authorities in several countries has shown that large-scale shutdowns can only be safely ended if a systematic ``test and trace" program \cite{singapore_covid19,salathe2020covid} is put in place to control the spread of the virus. This, in turn, is predicated on the widespread availability of mass diagnostic testing. However, most countries including the US are currently experiencing a scarcity \cite{ranney2020critical} of various medical resources including tests \cite{fair_allocation}.

One simple method to increase the effective testing capacity by testing pooled samples of a number of test subjects collectively instead of testing samples from each person individually. This idea of ``group testing" goes back many decades \cite{dorfman1943} and is based on the following intuition. If the rate of infection in the population is relatively low, statistically, most individual will test negative. With group testing, a single negative test result on a pooled sample immediately shows that all individuals in that pool are infection-free.

This potentially allows us to reduce the total number of tests {\it per subject} so the throughput of the existing testing infrastructure is increased \cite{hanel2020boosting} i.e. a much larger number of people can be tested compared to individual testing while keeping the number of tests the same.

Pooling does have its risks. The additional pre-processing required for preparing the pooled samples could affect the accuracy of the test because of possible degradation or contamination of the RNA. Pooling also requires dilution of the individual samples, and this in turn may increase the chances of a false negative result. However, pooling tests have been successfully used for diagnostic testing for infectious diseases in the past \cite{Taylor512,arnold_2013}. Preliminary studies on the Covid-19 virus also show that pooling samples \cite{Yelin2020} can be effective with existing tests.

The current testing bottlenecks in the Covid-19 crisis has led to a resurgence of interest in using group testing methods for Covid-19 diagnosis. Specifically, there have been recent studies \cite{Sinnott-Armstrong2020,Shani-Narkiss2020.04.06.20052159,poolingStanfordcommunity} into adapting pooling methods similar to \cite{dorfman1943} for Covid-19 testing. In \cite{zhu2020noisy}, the authors studied noisy group testing for virus detection. 

In this paper, we propose a different approach based on the compressed sensing theory \cite{OptimalRecovery} \cite{Candes05}\cite{donoho_compressed_2006}  for detection of viruses and antibodies using pooled sample testing. In compressed sensing, the measurement reading is not just a binary reading (`positive' or `negative') as in group testing, but instead the measurement reading of compressed sensing can be real-numbered quantification of the quantity of target DNA in the pooled sample. The traditional group testing methods such as \cite{dorfman1943} can be thought of as special cases of the more powerful compressed sensing framework proposed in this paper. This is because the measurement reading of group testing is a binary reduction of the real-numbered quantification of compressed sensing.   Through compressed sensing, it is possible to test $n$ persons for viruses by only using $O(k \log(n))$ samples,  where $k$ is the number of virus-infected persons. This is a significant reduction compared with testing each individual person, which would require $n$ testing. This can translate into an increase of test throughput in the order of $n/(k \log(n))$, which can be quite significant if the number of infected people is much smaller than the total population. 

Indeed, the real-numbered quantification from compressed sensing can greatly help speed up the testing of viruses and reduce the cost of testing, by taking advantage of the sparsity of virus infections in the population.  Compared with conventional group testing (including non-adaptive and adaptive group testing), compressed sensing has the following advantages:\\ (1) Compressed sensing uses real-numbered quantitative measurement results (quantification of target DNA etc. ) to infer virus infections or antibodies. These measurement readings contain 
more information about the collected samples than the binary readings of group testing.  This will make inference from compressed sensing measurements more robust against noises an outliers in the measurements, and require fewer tests. 

(2) Compressed sensing is known to require fewer measurements (or lower sample complexity) to infer virus infections than group testing. The sparsity $k$ that compressed sensing can handle for successful detection is allowed to grow linearly (proportionally) with $n$, while the recoverable sparsity $k$ is of the order $O(\sqrt{n})$ for non-adaptive group testing \cite{du1993combinatorial}. This will potentially translate higher testing throughput for compressed sensing than group testing. 

(3) The inference results from compressed sensing not only reveal which persons test positive or negative, but also reveal a quantitative evaluation of infections for the persons who test positive. For example, it can reveal the viral  loads (copies/ml) of persons who test positive. These quantitative results can help achieve better diagnosis and treatment of infected persons, and can also help study infectious power of viruses in different phases of infections.

There are broadly two types of tests for Covid-19: (a) serological tests that look for the presence of antibodies to the virus, or (b) swab tests that look for RNA from the live virus. While antibody tests have certain advantages e.g. can detect infections even after the subject has recovered, the most common tests currently used in the US and recommended by the CDC are swab tests. These tests use the Reverse Transcription Polymerase Chain Reaction (RT-PCR) process to selectively amplify DNA strands produced by viral RNA specific to the Covid-19 virus.

The RT-qPCR process which is considered the gold standard for the detection of mRNA consists of three distinct steps: (1) reverse transcription of RNA into cDNA, (2) selective amplification of a target DNA fragment using the Polymerase Chain Reaction (PCR), and (3) detection of the amplification product. While the simple ``end-point" version of PCR only allows binary detection (presence or absence) of a target RNA sequence, the real-time or quantitative version of the PCR process (qPCR) \cite{Gibson01101996} also allows the quantification of the RNA i.e. it produces an estimate of the quantity of the RNA material present in the sample \cite{nolan2006quantification}.

Some researchers \cite{Lamb2020} have proposed the Reverse Transcription Loop-Mediated Isothermal Amplification (RT-LAMP) as a potentially cheaper and faster alternative to RT-PCR for swab tests. While we focus on tests based on the RT-qPCR process, the methods proposed in this paper are also compatible with RT-LAMP \cite{Schmid-Burgk2020} and other DNA amplification methods.

In this paper, we propose to use compressed sensing to detect viruses and antibodies of COVID-19.  Considering the physical and complexity constraints of pooling for compressed sensing, we identify sparse bipartite graph based measurement matrices for compressed sensing applied to this purpose. In particular, we propose to use expander graph based measurement matrices \cite{Weiyuexpander2007} for pooling or measurement designs.   

As mentioned above, group testing has a long history of being used to detect pathogens, tracing back to World War II, and it has also recently been applied to testing COVID-19 viruses \cite{Sinnott-Armstrong2020,Shani-Narkiss2020.04.06.20052159}. To the best of our knowledge, this work might be the first to develop compressed sensing techniques for detecting viruses using qPCR and other tools, especially when applied to COVID-19 viruses \footnote{The authors of this paper conceived the idea of performing group testing and compressed sensing for COVID-19 virus detection during the early outbreak in China,  and the pandemic status of COVID-19 has motivated us to carry out this research. }. On a related but different subject, we note that compressed sensing was proposed in \cite{ShentalGenetic2010} to study human genetics, and used to identity people with rare alleles (``allele is one of two or more alternative forms of a gene that arise by mutation and are found at the same place on a chromosome.'').  
\section{Compressed Sensing for High-Throughput Virus Detection: System Model and Problem Formulation}

In this section, we describe the system architecture of using compressed sensing to speed up the testing of COVID-19 viruses or antibodies, including sensing matrix design and decoding algorithm design. We will focus on developing such systems  using Polymerase Chain Reaction (PCR) machines, especially real-time PCR (quantitative PCR, qPCR or RT-PCR) machines,  to test the viruses, though the concepts and ideas introduced in this paper extend to testing viruses using other technologies or platforms and also to testing antibodies.  (We note that in the literature, there are inconsistencies about the meanings of ``RT-PCR'', which are  used as abbreviations for both reverse transcription PCR  and real-time PCR.)
We start by introducing some background knowledge on the real-time quantitative PCR \cite{PCRHandbook}.

\begin{figure}

			\includegraphics[width=\linewidth]{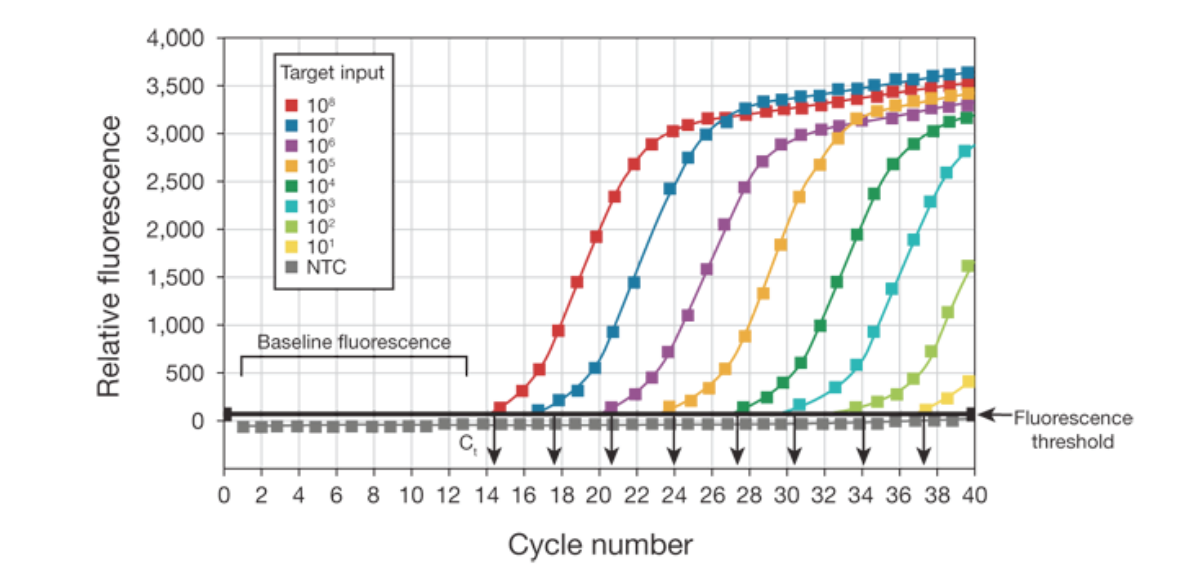}
			
		\caption{Amplication plots of real-time polymerase chain reaction (PCR) taken from \cite{PCRHandbook}.  According to \cite{PCRHandbook}, this figure is about ``Relative fluorescence vs. cycle number.''     ``Amplification plots are created when the fluorescent signal from each sample is plotted against
cycle number; therefore, amplification plots represent the accumulation of product over the duration of the real-time PCR experiment. The samples
used to create the plots in this figure are a dilution series of the target DNA sequence.'' \cite{PCRHandbook}  }\label{Fig:RTPCR}
	\end{figure}

The polymerase chain reaction (PCR) is one of the most powerful and widely used technologies in molecular biology to detect and quantify specific sequences within a DNA or cDNA template. Using PCR, specific sequences within a DNA or cDNA template can be copied, or “amplified”, to thousands or to a million times
using sequence-specific oligonucleotides, heat-stable DNA polymerase, and thermal cycling \cite{kralik_basic_2017}. PCR theoretically amplifies DNA exponentially, doubling the number of target molecules with each amplification cycle.

To address the need of robust quantification of DNA,  real-time polymerase chain reaction (real-time PCR) was developed based on the polymerase chain reaction (PCR).  Real-time PCR is carried out in a thermal cycler (providing temperature conditions for each cycle of reactions),  but with the capacity to illuminate each sample with a beam of light and detect the fluorescence emitted by the excited fluorophore \cite{PCRHandbook}.

In traditional (endpoint) PCR, detection and quantification of the amplified sequence are performed at the end of the reaction after the last PCR cycle. In real-time quantitative PCR, PCR product (the amplified sequences) is measured at each PCR cycle. Namely, Real-time PCR can monitor the amplification of a targeted DNA module in the PCR in real time. By monitoring reactions during the exponential amplification phase of the reaction, users can determine the initial quantity of the target with great precision. The working physical principle of the RT-PCR is that it detects amplification of DNA in real time by the use of fluorescent reporter. The fluorescent reporter signal strength is directly proportional to the number of amplified DNA molecules.

Real-time PCR commonly relies on plotting fluorescence against the number of cycles on a logarithmic scale to perform DNA quantification.  During the exponential amplification phase, the quantity of the target DNA template (amplicon) doubles every cycle. A threshold for detection of DNA-based fluorescence is set 3–5 times of the standard deviation of the signal noise above background. The number of cycles at which the fluorescence exceeds the threshold is called the threshold cycle ($C_t$) or, quantification cycle ($C_q$). One can then use this threshold cycle $C_t$ to determine the quantity of target DNA in the sample. In ideal cases, if the threshold cycle of a DNA sample $A$ precedes that of another sample $B$ by $N$ cycles, then this DNA sample $A$ contains $2^N$ times more target DNAs than DNA sample $B$ at the beginning of the reaction.  In practice, people often use the standard curve method for real-time PCR to determine the relation between threshold cycle $C_t$ and target quantity.

\subsection{Compressed Sensing System for High-Throughput Rapid Testing}

In this subsection, we propose and describe a compressed sensing system to perform high-throughput rapid testing of COVID-19 and antibodies. We remark that this system also applies to testing of other types of viruses or antibodies. 

 Suppose that we have collected $n$ samples of $n$ persons, and we would like to test how many among them have viruses and what quantity of viruses they have. (It is also possible that we can collect more than 1 sample from a person, but for simplicity of presentations, we stick with 1 sample per person.)   We use a non-negative vector $x \in \mathbb{R}^n$ to denote the quantities of COVID-19 viruses in the samples of these $n$ persons, where $x_i$, the $i$-th element of $x$, corresponds to the quantity of target DNA in the sample of the $i$-th person, and $\mathbb{R}$ is the set of real numbers. If the $i$-th person is not infected or has no COVID-19 virus, $x_i=0$ or very close to $0$; if instead the $i$-th person is infected, $x_i>0$. If there are $k$ ($k$ can be small compared with $n$) people affected among these $n$ persons, $x$ will have $k$ positive elements, and the rest of its elements are zero.  This leads to a sparse $x$, and we call such a vector $k$-sparse vector, meaning it only has $k$ nonzero elements. 
 When the vector $x$ is sparse, compressed sensing theories offer to greatly reduce the number of testings that need to be done to accurately infer $x$ \cite{Candes05} \cite{donoho_compressed_2006}. This implies high-throughput, fast and low-cost testing for detecting viruses. The basic idea of compressed sensing is to observe mixed or pooled samples of elements of $x$ through a wide measurement matrix (as introduced below). Compared with group testing, compressed sensing can correctly infer the real-numbered values of $x$ (which will be useful for research of different phases of infections, better diagnosis, treatment of infected persons), requires fewer testing to detect positive cases, and is more robust against noisy observations.

 We then design mixing matrix $E$ of dimension $m\times n$, where $m$ can be significantly smaller than $n$.  In fact, $m$ is the number of tests we will eventually need to run to detect viruses, and often we have $m \ll n$, thus making the tests more efficient and increasing the throughout of the tests.  We let each element of $E$ be either $0$ or $1$.  We denote the element of $E$ in the $i$-th row and $j$-th column as $E_{i, j}$:
 
 \begin{equation}
  E_{i,j} =
    \begin{cases}
      1 & \text{if sample $j$ participates in testing $i$,}\\
      0 & \text{otherwise.}
    \end{cases}       
\end{equation}

Namely, if $E_{i,j}=1$, where $1\leq i \leq m$ and $1\leq j \leq n$, (part of) the $j$-th person's biological sample will be mixed with samples from other persons, and we will perform PCR (or other testing technologies) over this mixed sample in the $i$-th test. Otherwise, the $i$-th test will not involve the $j$-th person. The sample of the $j$-th person can be involved in multiple testings, the number of which is equal to the number of '1's in the $j$-th column of $E$.

Since a person's sample is involved in multiple testings, we need to allocate a portion of that person's sample for each of the involved testings of that person.  Thus for each  $j$, $1\leq j\leq n$,  we associate the $j$-th person's sample with an ''allocation" vector $w_j \in \mathbb{R}^m$, whose elements are nonnegative and $\|w_j\|_1 \leq 1$ (the summation of $w_j$'s elements are no more than 1).  For example, if the $i$-element of $w_j$ is 0.2, it means that 20 percent of the sample from the $j$ person participates in the $i$-th testing. 

Using $w_j$'s, we can form an allocation matrix $W$ as 
$$W=[w_1, w_2,..., w_n].$$

We define the actual measurement matrix $A$ of dimension $m\times n$ as
$$A=E\odot W,$$
where $\odot$ means elementwise multiplication. 

Then the generalized compressed sensing testing result vector $y\in\mathbb{R}^n $ is given by 
$$y=f(A \times x)+v+e,$$
where each element of $y$ represents the measurement results of the DNA quantity in a single test (as can be computed by looking at the threshold cycle $C_t$'s value), $f(\cdot): \mathbb{R}^m\rightarrow \mathbb{R}^m$ is a functional modeling non-linearity and randomness associated with the measurement process, $v$ is a random noise vector and $e$ can be a vector containing potential outliers. 

As a special case, in an ideal real-time PCR, we can have $f(Ax)=Ax$. However, this formulation is very general, and can be used to model other types of non-linearity or randomness in testing. For example, for a traditional end-point PCR or if we only use the real-time PCR to see whether viruses exist, the functional $f(\cdot)$ can output a vector of  `true' or `false' depending on whether the quantity of DNA samples is above a certain significance threshold. In this paper, we focus on the RT-PCR, and assume it is ideal in the sense that the quantity of DNA sample inferred from its readings is $f(Ax)=Ax$. Compared with group testing, in our compressed sensing systems, the output $y$ can work with real numbers or  other general formats such as the whole amplication plot of qPCR, and can glean more information from each test (or measurement) than binary information. Since compressed sensing can retain more information about the vector $x$, in general fewer tests are needed for inferring $x$ or the support of $x$. For example, compressed sensing can only use $m=O(k \log (\frac{n}{k}))$ tests to fully recover $x$, while group testing needs $m=O(k^2 log (\frac{n}{k}))$ tests \cite{du1993combinatorial}.

\subsection{Design of Measurement Matrix $A$}
To achieve robust and rapid testing, we design the matrix $A$ in the following ways. 

(1) Recall that we have the measurement matrix $A=E \odot W$, where $E$ is the mixing matrix and $W$ is the allocation matrix. The matrix $E$ is a 0-1 matrix with '0' or '1' elements. The number of '1''s in the matrix $E$ should be small, thus making matrix $E$ a sparse matrix.  This is because we would like the number of '1's in $E$ to be as small as possible in order to minimize the complexity of mixing samples from different persons, and minimize the probability of mistakes in mixing.  For each column of $E$, we also consider constraining the number of `1''s. This is because we do not want to dilute the quantity of the $j$-th person's sample too much by distributing it to too many tests. If it is distributed to too many tests, the quantity from the $j$-th person for each individual test can be too little for going above the detection threshold of the PCR machines.

All these physical constraints and considerations motivate us to propose using sparse bipartite graph measurement matrices for the design of $E$ and $A$.  In particular, we propose to use the expander-graph based compressed sensing, which was proposed for general compressed sensing by one author of this paper \cite{Weiyuexpander2007}\cite{Sina}.
The expander graph based measurement matrix is a 0-1 matrix derived from expander bipartite graphs. It comes with efficient decoding algorithms and provable performance guarantees for testing. Moreover, the number of '1's in each column can be upper bounded for the expander graph based matrices, which complies with the physical constraint that a person's sample cannot be distributed to too many samples.

(2) We have the freedom of designing the allocation matrix, but, for simplest presentations, in this paper, we can choose the simplest allocation design of evenly dividing the sample into the measurements involved. Namely,  $A$ will be obtained by dividing each column of $E$ by the total number of `1's in that column. It is entirely possible to use other allocation matrices for better performance or more efficient decoding.


(4) Considering physical and operational constraints, matrix $A$ cannot be too wide and too tall at the same time.

\subsection{Detection (Decoding) algorithms from compressed mixed measurements}

From the measurement result $y$, one can infer the quantity of DNA sample (or viruses) associated with each person. Due to the extensive developments of compressed sensing \cite{Candes05}\cite{donoho_compressed_2006} over the last two decades, there are many decoding algorithms to infer $x$ from $y$, such as basis pursuit ($\ell_1$ minimization), LASSO, message passing style algorithms \cite{Weiyuexpander2007} \cite{Donohomessagepassing}, and greedy algorithms such as orthogonal matching pursuits.  One can potentially choose any of these algorithms to do the decoding. We also notice that the signal $x$ is nonnegative, which can be used to boost the efficiency of compressed sensing \cite{AminNonnegativeExpander}.

 However, for detecting viruses or antibodies, we still need to choose or develop fast and robust decoding algorithms in this particular application. The reason is that many of the aforementioned algorithms have performance guarantees or good empirical performance when the dimensions of $A$ are very large or $m$ is asymptotically proportional to $n$ when $n$ goes to infinity. This is not the case for compressed sensing for virus detection, since we have a measurement matrix of finite and possible very limited sizes.  Some of these algorithms can experience severe performance degradation because of size limitations of $A$.

 Because of the limited sizes of matrix $A$, and to reduce the false positive rate and false negative rates of the testing, we can start with the following two algorithms, and the message passing style iterative algorithms for expander graphs \cite{Weiyuexpander2007}: \\
 
 (1) $\ell_0$ minimization.  
 
 This is equivalent to exhaustive search over all the possible sets of $k$ persons with viruses，and then solve for $x$ using an overdetermined systems for each of these sets using $y$. Formally, if there is no noise in the observation, we are solving
 
\setlength{\mathindent}{15pt}
\par\noindent
\small
\begin{align}
\label{L0_min}
          & \text{minimize} \;\; \| x \|_{0} \nonumber\\
          & \text{subject to} \;\; y = Ax,\\
          &\;\;\;\;\;\;~~~~~~~~~~x\geq 0.
\end{align}
\normalsize
where $\|x\|_0$ is the number of non-zero elements in vector $x$. The $\ell_0$ minimization is an NP-hard problem. But the exhaustive search or its modifications might be good choice for this application, since it gives great performance in minimizing false positive rate and false negative rate.  Since the problem of size of this application may not be big due to physical constraints, it can be computationally feasible.

(2) $\ell_1$ minimization.

To reduce the computational complexity, we can often relax (\ref{L0_min}) to its closest convex approximation - the $\ell_1$ minimization problem:
\par\noindent
\small
\begin{align}
\label{L1_min}
         & \text{minimize} \;\; \| x \|_{1}  \nonumber\\
         & \text{subject to} \;\; y = Ax,\\
         & \;\;\;\;\;\;\; ~~~~~~~~~x\geq 0,
\end{align}
\normalsize
where $\|x\|_1$ is the sum of the absolute values of all the elements in $x$.

After solving for the vector $x$, we can set a threshold $\tau>0$ such that if $x_j\leq \tau$, we declare the test is positive for the $j$-th person; otherwise, we declare the testing result as negative.

It has been shown that the optimal solution of $\ell_0$ minimization can be obtained by solving $\ell_1$ minimization under certain conditions (e.g. Restricted Isometry Property or RIP)  \cite{Candes05} \cite{candes_robust_2006} \cite{donoho_compressed_2006}\cite{Donoho06}\cite{DTCISS06}.
A necessary and sufficient condition under which a vector $x$ with no more than $k$ nonzero elements can be uniquely obtained via $\ell_1$ minimization is Null Space Condition (NSC), for example, see \cite{devore2,PreciseStability11}. While the RIP condition and NSC condition are normally satisfied for large-dimension matrix $A$, there are algorithms which can precisely verify the null space condition for small-size problems, which will be especially useful for designing optimal pooling strategies or the compressed sensing matrices for detection of viruses. \cite{Juditsky}\cite{TestingNull} \cite{ChoTesting}.
\section{Numerical Experiments}
\label{sec_result}
In the experiments, we consider two types of binary pooling matrix: Bernoulli random matrix where each entry of the matrix is `0' with probability 0.5, and is `1' with probability 0.5, and measurement matrix obtained from an expander graph \cite{Weiyuexpander2007} where each column has a fixed number of ones.  Experimenting with random Bernoulli pooling matrices can show the typical performance of such pooling matrices. In practice, one needs to work with deterministic pooling matrices.  To design a deterministic matrices, one can use algorithms in \cite{ChoTesting} to precisely verify the performance guarantee of a randomly generated matrix for virus testing. After the verification, we can then use it as a deterministic pooling matrix in practice. 

For these two types of binary pooling matrix, we consider two  different values for the number of people tested, i.e., $n=120$ and $n=60$. For each of the two values of length $n$, we recover the value of $x$ with different sparsity (sparsity is the number of people infected in this group of people), i.e., $k=3$ and $k=5$. In the experiments, we set random $k$ entries of the signal of length $n$ to be random numbers within $[5,10]$, while the other entries are set to be positive numbers close to 0.

\begin{figure}
		\includegraphics[width=0.5\textwidth, height=0.3\textwidth]{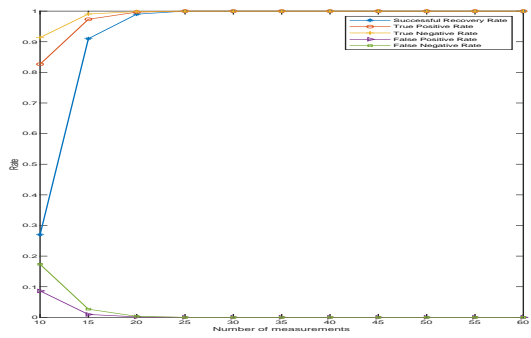}
		\caption{$n=60, k=3$. Binary measurement matrix with entries i.i.d. according to Bernoulli distribution.}
		\label{Fig:Bernoulli1}
\end{figure}

\begin{figure}
		\includegraphics[width=0.5\textwidth, height=0.3\textwidth]{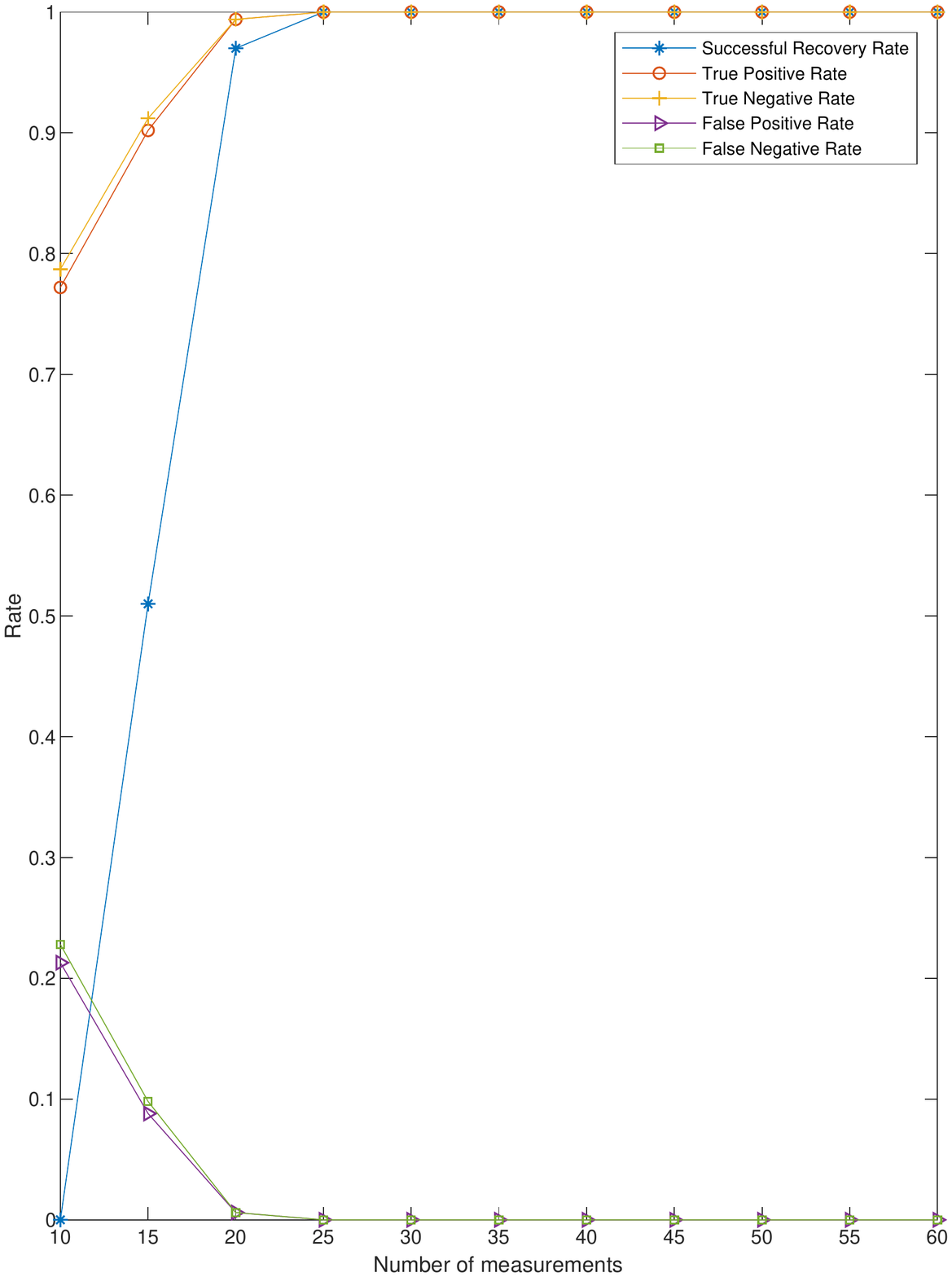}
		\caption{$n=60,k=5$. Binary measurement matrix with entries i.i.d. according to Bernoulli distribution.}
\label{Fig:Bernoulli2}
\end{figure}

\begin{figure}
	\centering
	\includegraphics[width=0.5\textwidth, height=0.3\textwidth]{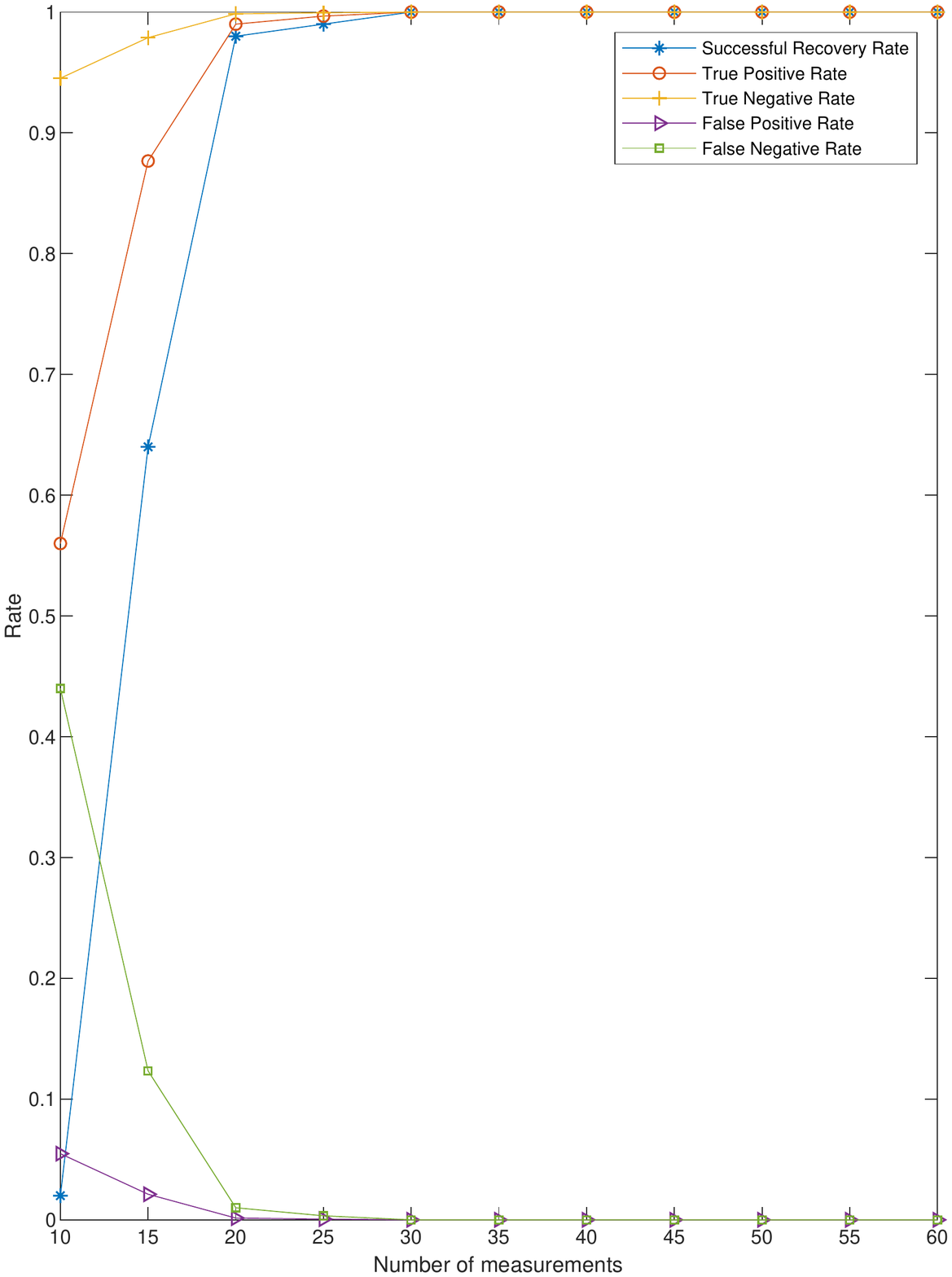}
	\caption{$n=120, k=3$.Binary measurement matrix with entries i.i.d. according to Bernoulli distribution.}
	\label{Fig:Bernoulli3}
\end{figure}

\begin{figure}
	\centering
	\includegraphics [width=0.5\textwidth, height=0.3\textwidth] {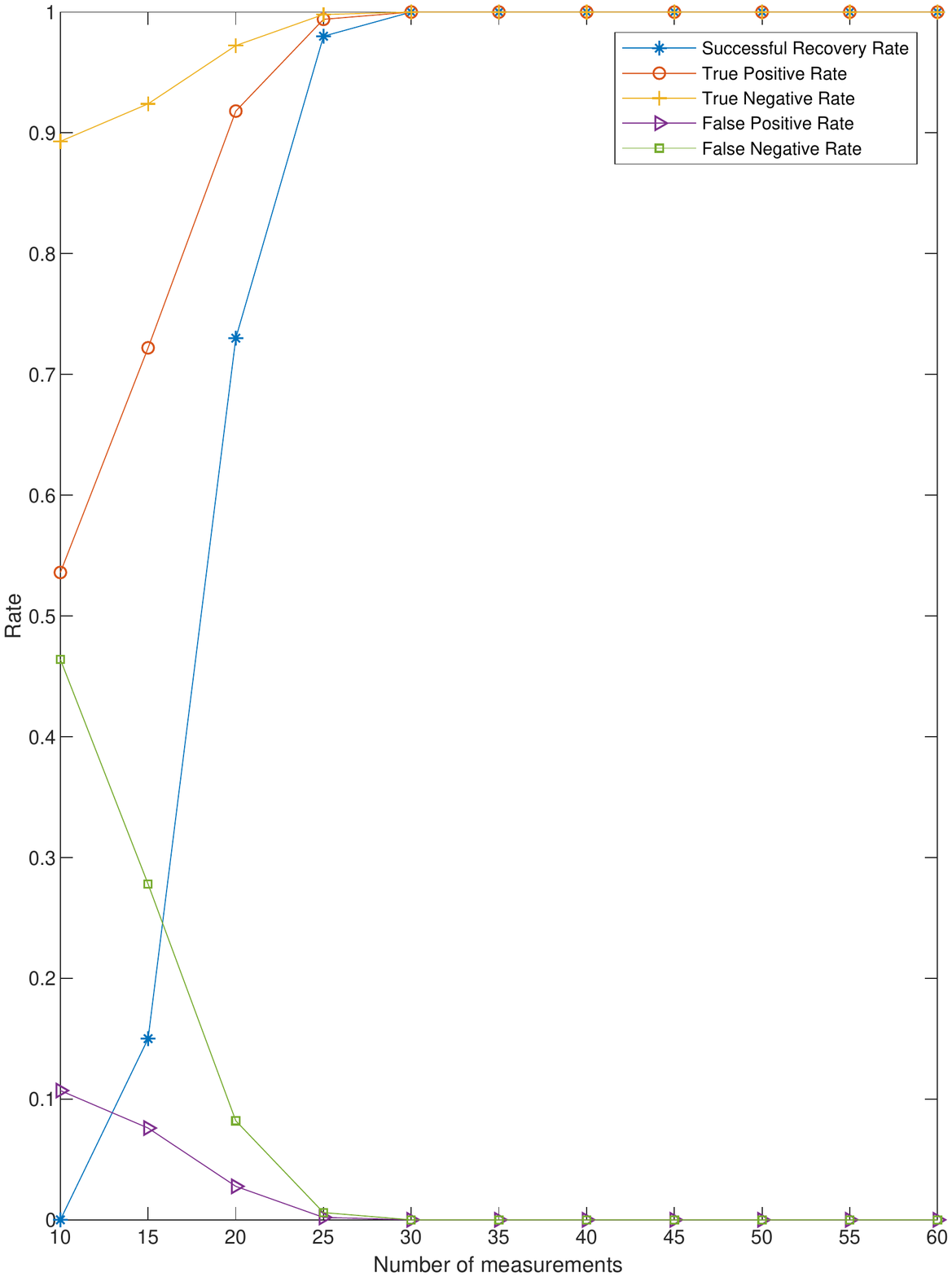}
	\caption{$n=120,k=5$. Binary measurement matrix with entries i.i.d. according to Bernoulli distribution.}
	\label{Fig:Bernoulli4}
\end{figure}

\begin{figure}
	\centering
		\includegraphics[width=0.5\textwidth, height=0.3\textwidth] {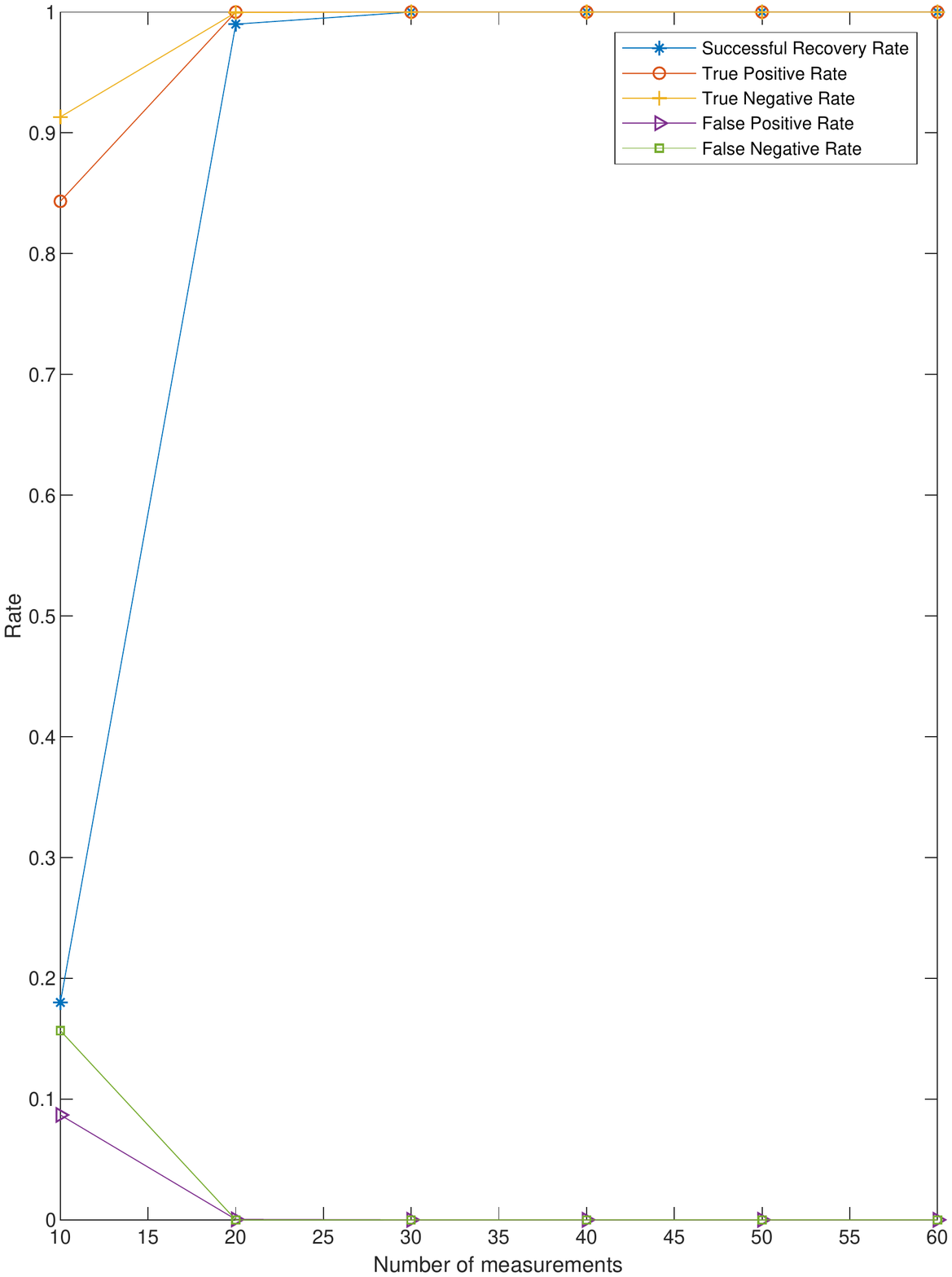}
		\caption{$n=60, k=3$. Expander measurement matrix with 5 '1's in each column.}
		\label{Fig:Expander1}
	\end{figure}

\begin{figure}
		
		\includegraphics[width=0.5\textwidth, height=0.3\textwidth]{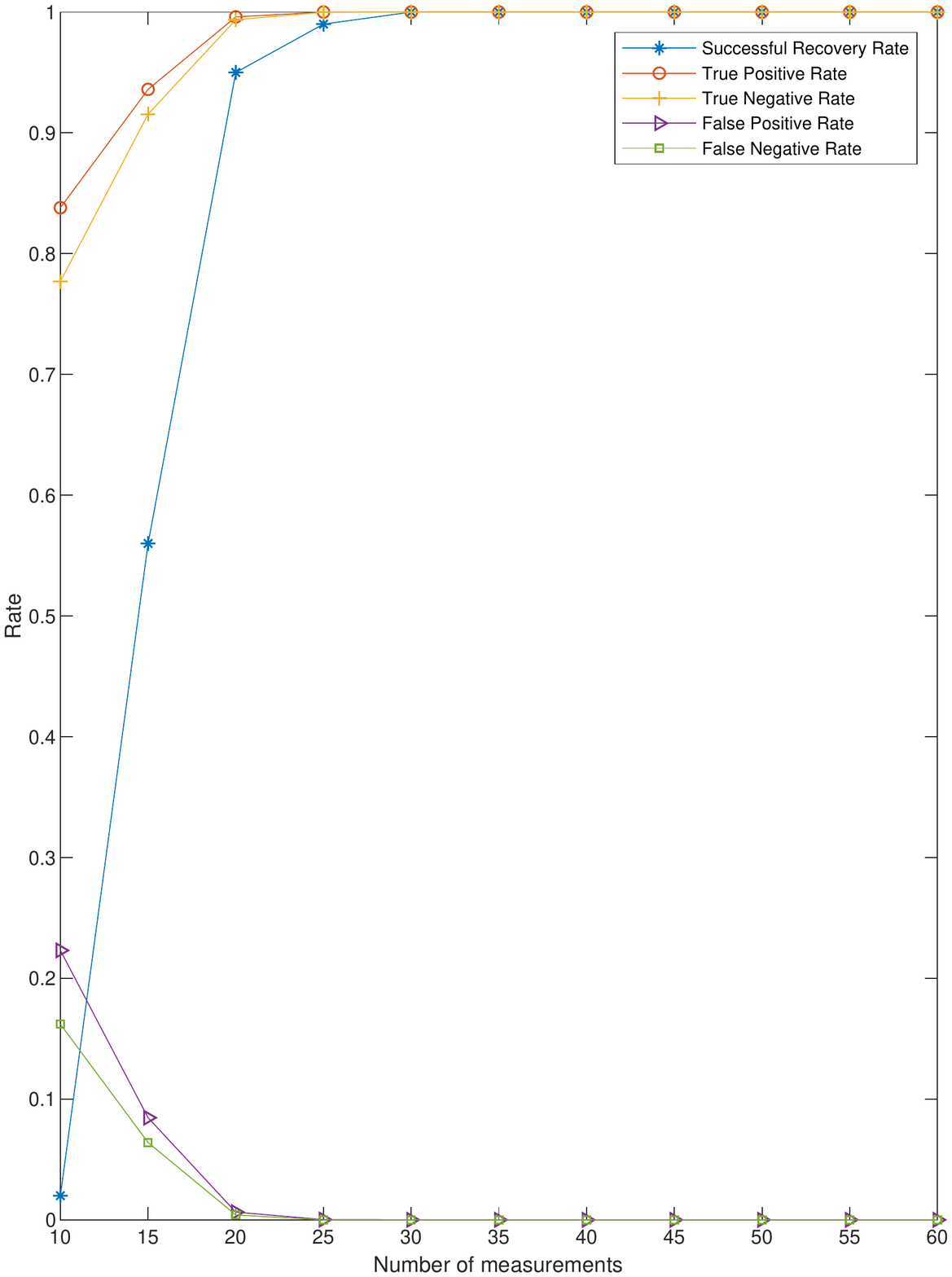}
		\caption{$n=60,k=5$. Expander measurement matrix with 5 '1's in each column.}
		\label{Fig:Expander2}
	\end{figure}

\begin{figure}
		\centering
		\includegraphics[width=0.5\textwidth, height=0.3\textwidth]{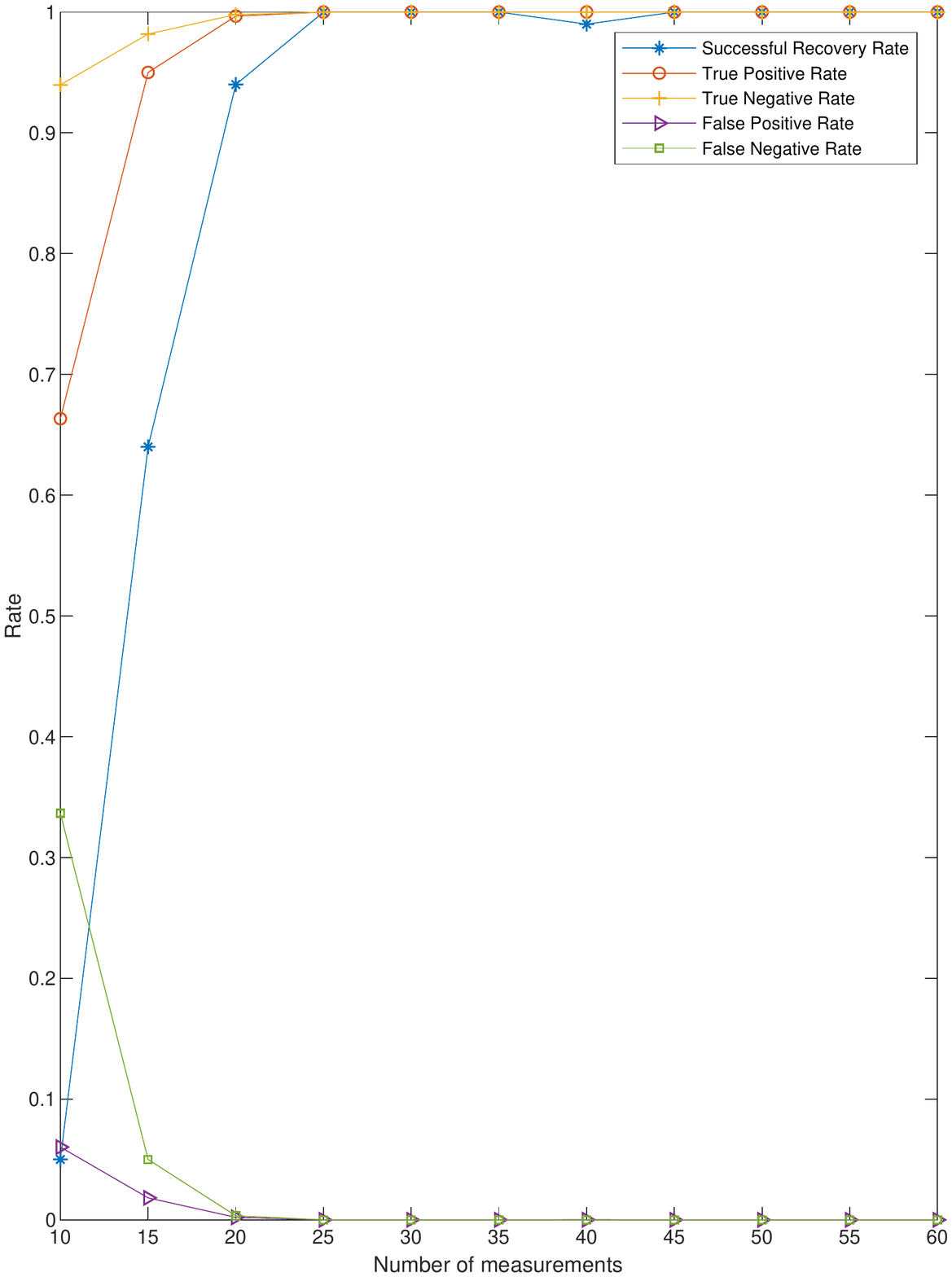}
		\caption{$n=120, k=3$. Expander measurement matrix with 5 '1's in each column.}
		\label{Fig:Expander3}
	\end{figure}

\begin{figure}
		\centering
		\includegraphics[width=0.5\textwidth, height=0.3\textwidth] {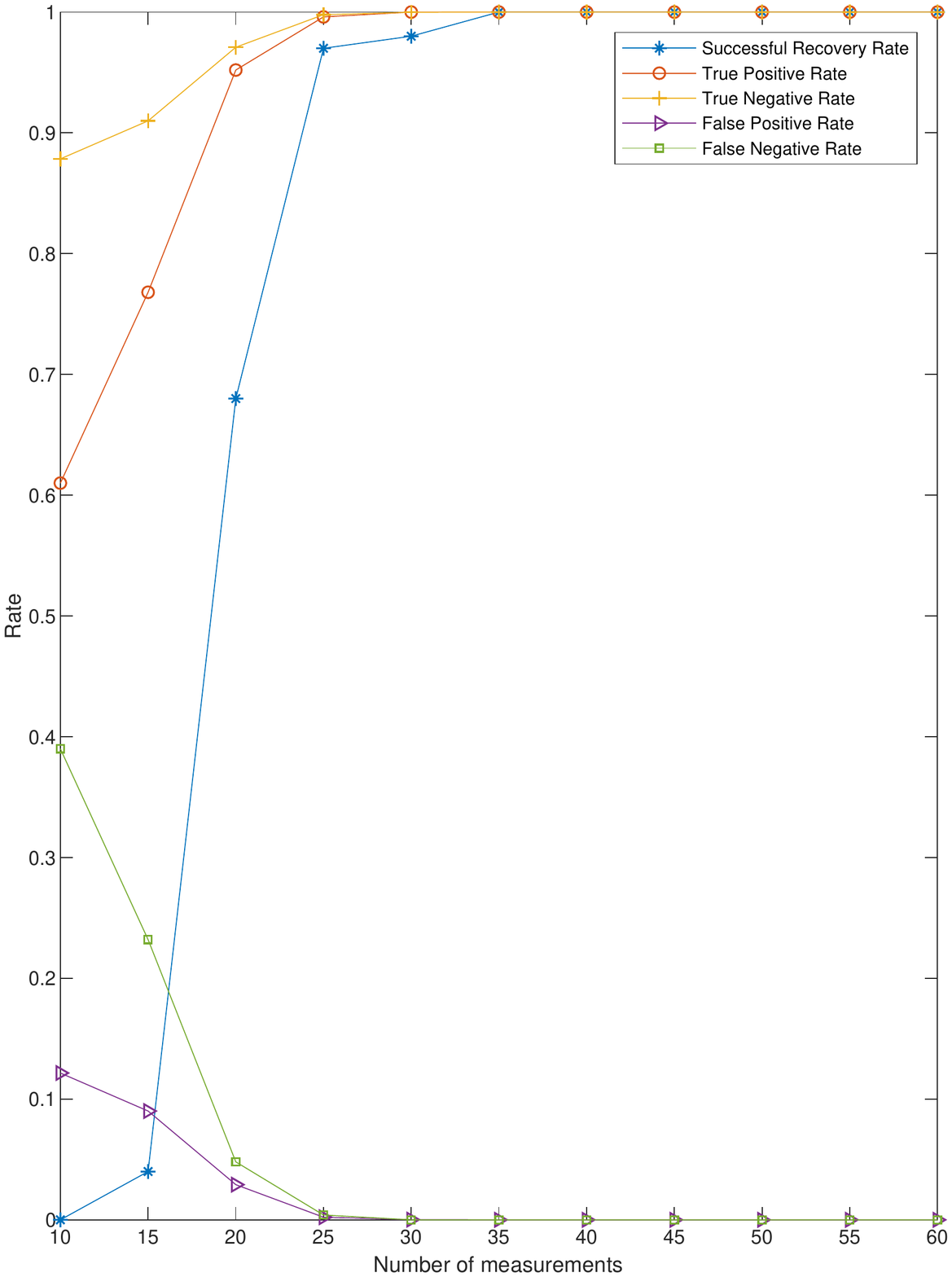}
		\caption{$n=120,k=5$. Expander measurement matrix with 5 '1's in each column.}
		\label{Fig:Expander4}
	\end{figure}

When $n=60$, for each $k$ and measurement matrix type, we take different measurements $m$=$10$,$15$,$20$,..., and $60$. For each possible $m$, we run 100 trials to evaluate the successful recovery rate via solving 
\begin{align}
\min_{x\in\mbR^n} \|x\|_1, {\rm s.t.\ }Ax=y, x\geq 0,
\end{align}
where $A\in\mbR^{m\times n}$ is the measurement matrix, $x$ is the signal to be recovered, and $y\in\mbR^m$ is the measurement vector.  After a signal is decoded, we use a thresholding technique to identify the persons with viruses. For each trial, we set a threshold $\tau=0.5$. The signal entry will determined to be `positive' with viruses,  if the recovered value is at least $\tau$, and `negative' if it is less than $\tau$. We then calculate the true positive rate (TPR),true negative rate (TNR), false positive rate (FPR), and false negative rate (DNR). We also consider the recovery success rate: if the  reconstruction error (the Euclidean distance between the true signal $x$ and the recovered signal $\hat{x}$) is smaller than $10^{-3}$, we count the recovery as a success. The numerical results are shown in Figure \ref{Fig:Bernoulli1} to Figure 
\ref{Fig:Bernoulli4} for Bernoulli measurement matrices. Numerical results are shown in Figures \ref{Fig:Expander1} to \ref{Fig:Expander4} for expander graph based measurement matrices. As we can see from these figures,  $n=60$, we only need around $m=20$ tests to achieve very low false negative rates and false positive rates, which means that we can increase the throughput of virus testing by $\frac{n}{m}\approx 3 $ times.  For $n=120$, we also need around $m=20$ tests to achieve low false negative and false positive rates, which translates to around $\frac{n}{m}\approx 6 $ times increase in test throughput. 
For $k=2$ and $n=200$, when we use expander graph based pooling matrix with 5 `1's in each column, we can already achieve a near zero false positive and false negative rates when $m=20$. This translates to a $\frac{200}{20}$ folds of speedup in test throughput.

	~ 
	
	~ 

\begin{figure}
		\centering
		\includegraphics[width=0.5\textwidth, height=0.3\textwidth] {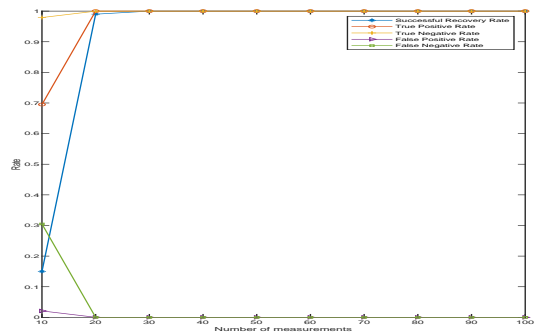}
		\caption{$n=200,k=2$. Expander measurement matrix with 5 '1's in each column. Noisy measurements.}
		\label{Fig:ExpanderNoisy10fold}
	\end{figure}

\begin{figure}
		\centering
		\includegraphics[width=0.5\textwidth, height=0.3\textwidth]{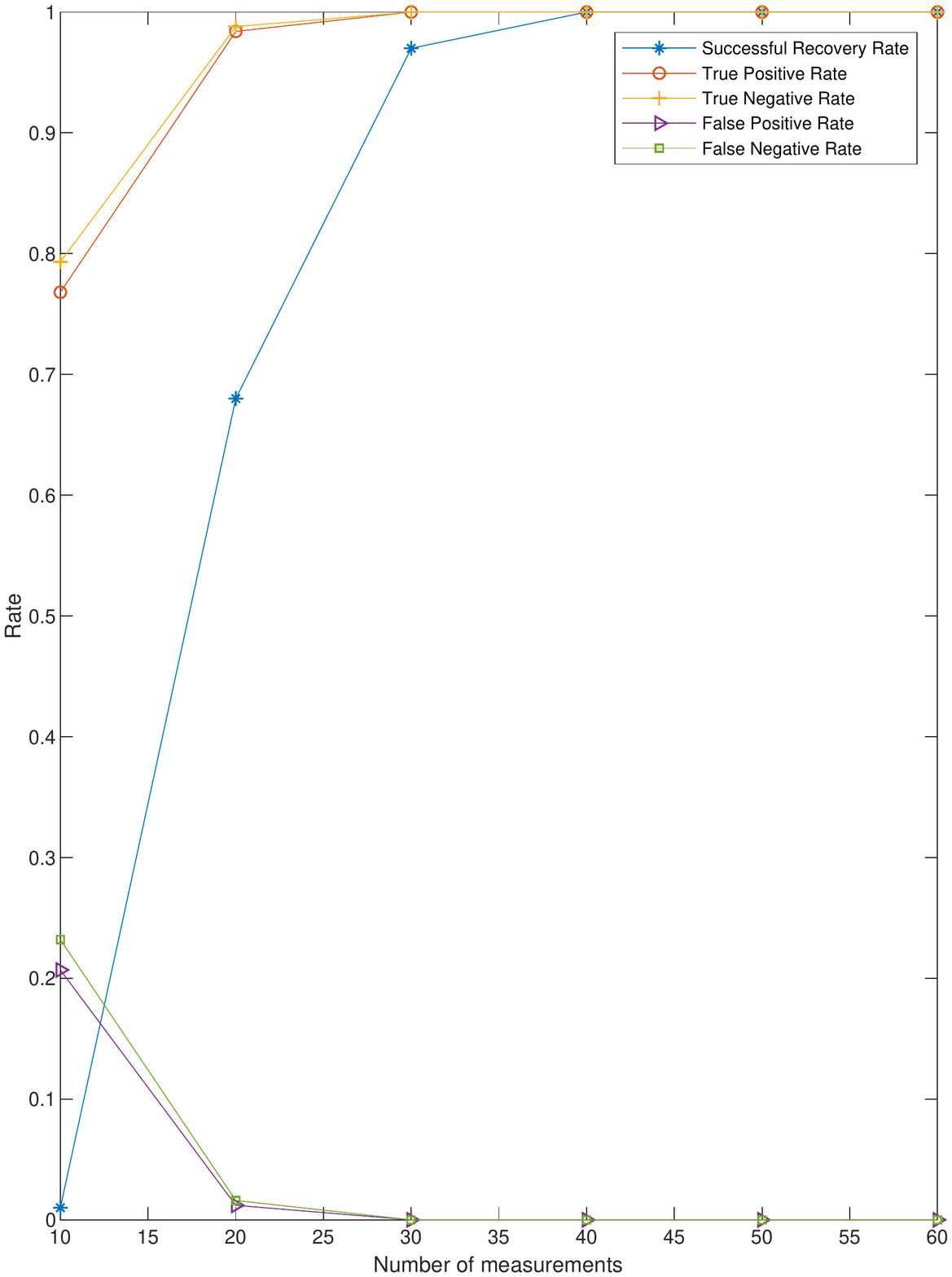}
		\caption{$n=60,k=5$. Binary measurement matrix with entries i.i.d. according to Bernoulli distribution. Noisy measurements.}
		\label{Fig:NoisyBernoulli2}
	\end{figure}

\begin{figure} 
		\centering
		\includegraphics[width=0.5\textwidth, height=0.3\textwidth]{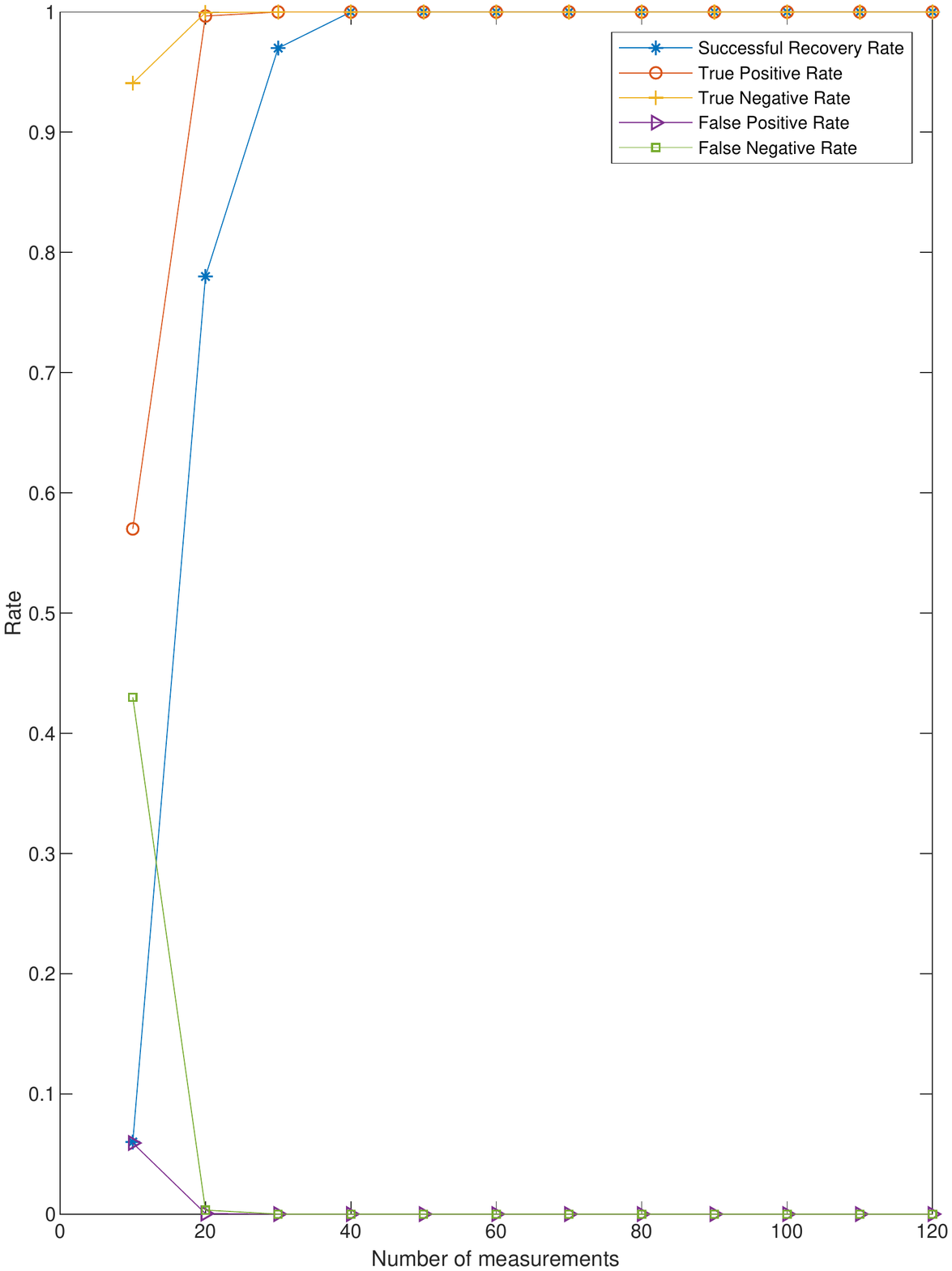}
		\caption{$n=120, k=3$. Binary measurement matrix with entries i.i.d. according to Bernoulli distribution. Noisy measurements.}
		\label{Fig:NoisyBernoulli3}
	\end{figure}

\begin{figure}		
	\centering
		\includegraphics [width=0.5\textwidth, height=0.3\textwidth] {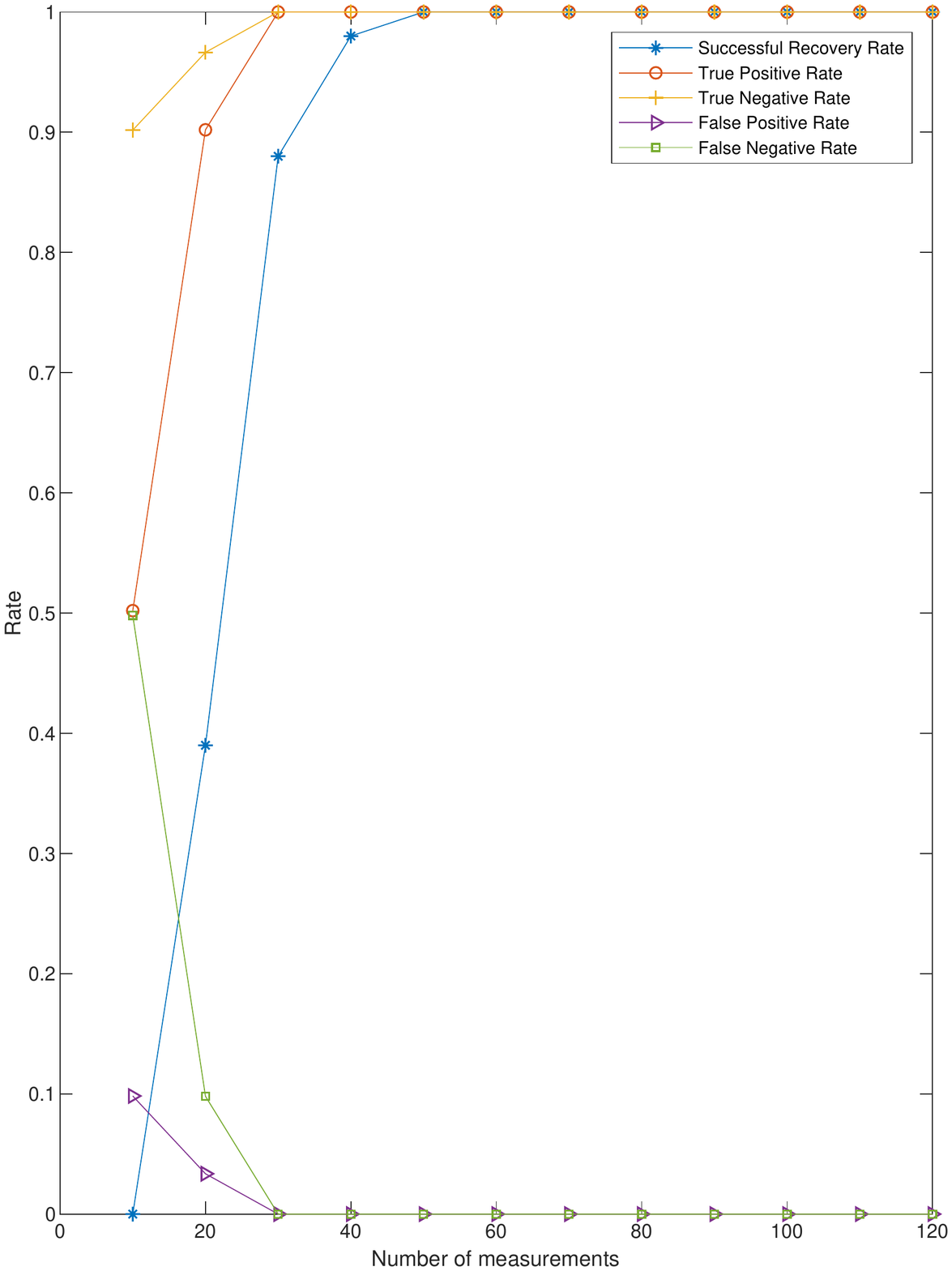}
		\caption{$n=120,k=5$. Binary measurement matrix with entries i.i.d. according to Bernoulli distribution. Noisy measurements.}
		\label{Fig:NoisyBernoulli4}
	\end{figure}

\begin{figure}
	\centering
		\includegraphics[width=0.5\textwidth, height=0.3\textwidth]{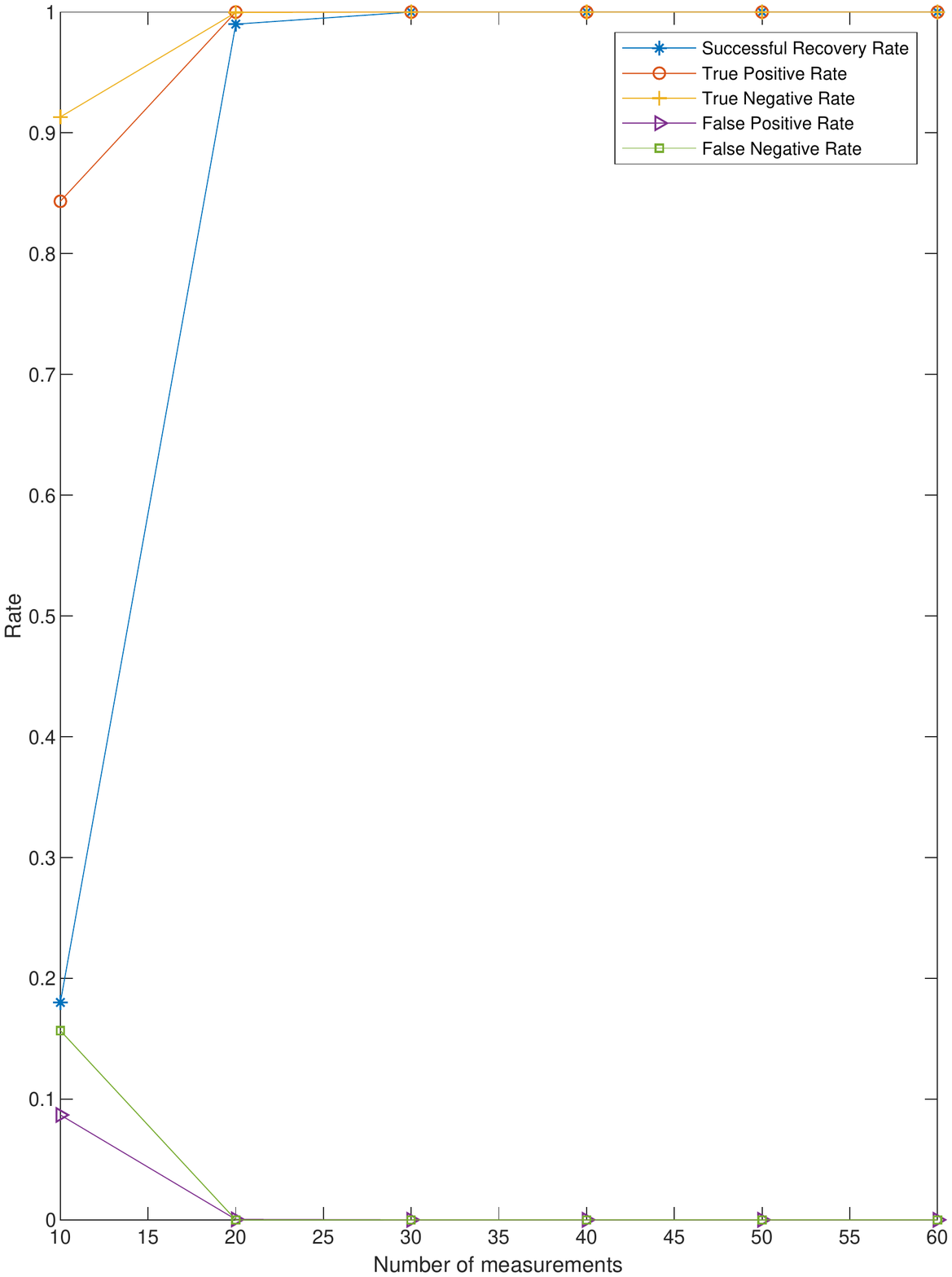}
		\caption{$n=60, k=3$. Expander measurement matrix with 5 '1's in each column. Noisy measurements.}
		\label{Fig:NoisyExpander1}
	\end{figure}

\begin{figure}
		\centering
		\includegraphics[width=0.5\textwidth, height=0.3\textwidth] {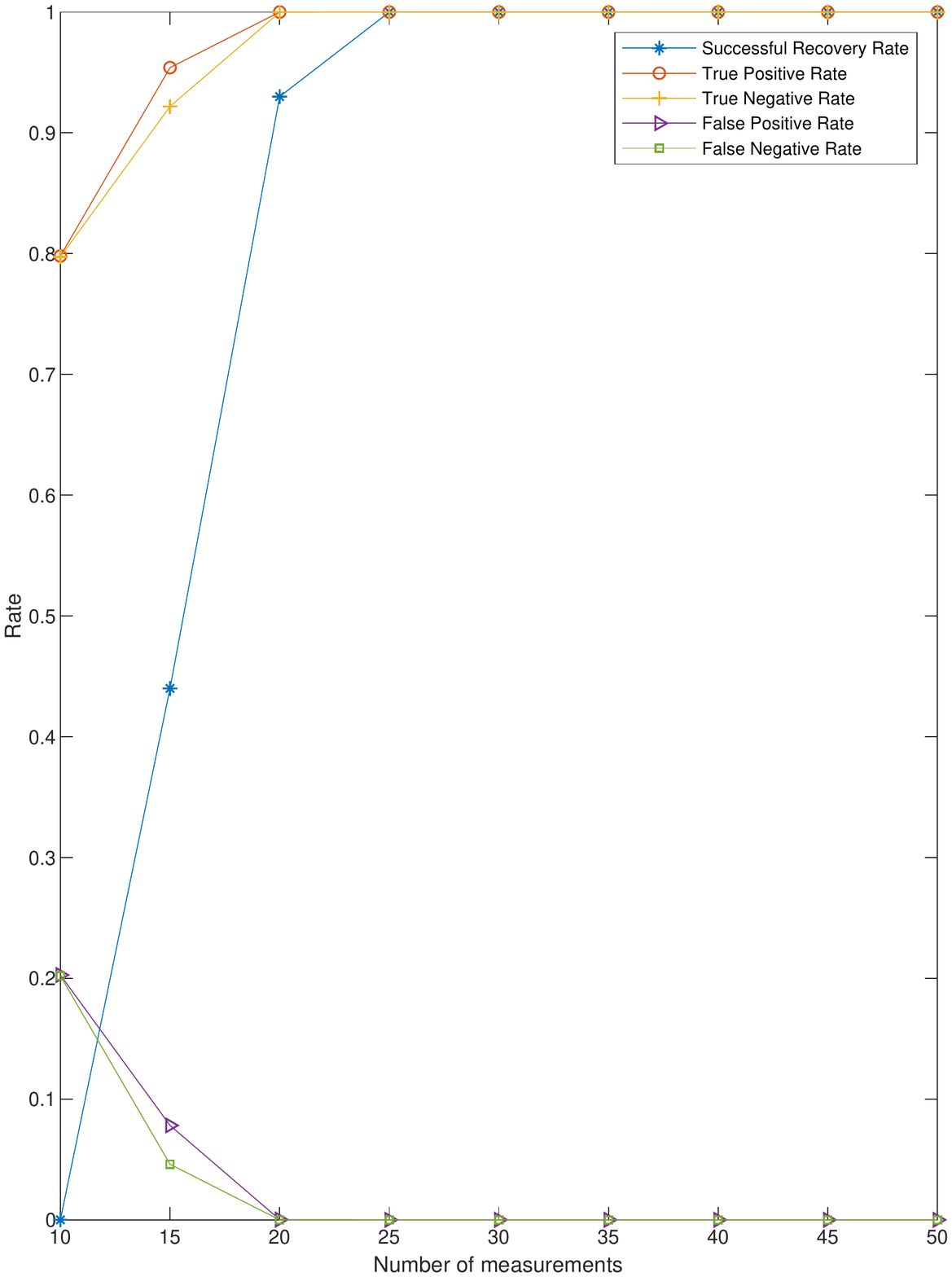}
		\caption{$n=60,k=5$. Expander measurement matrix with 5 '1's in each column.Noisy measurements.}
		\label{Fig:NoisyExpander2}
	\end{figure}

\begin{figure}
		\centering
		\includegraphics[width=0.5\textwidth, height=0.3\textwidth]{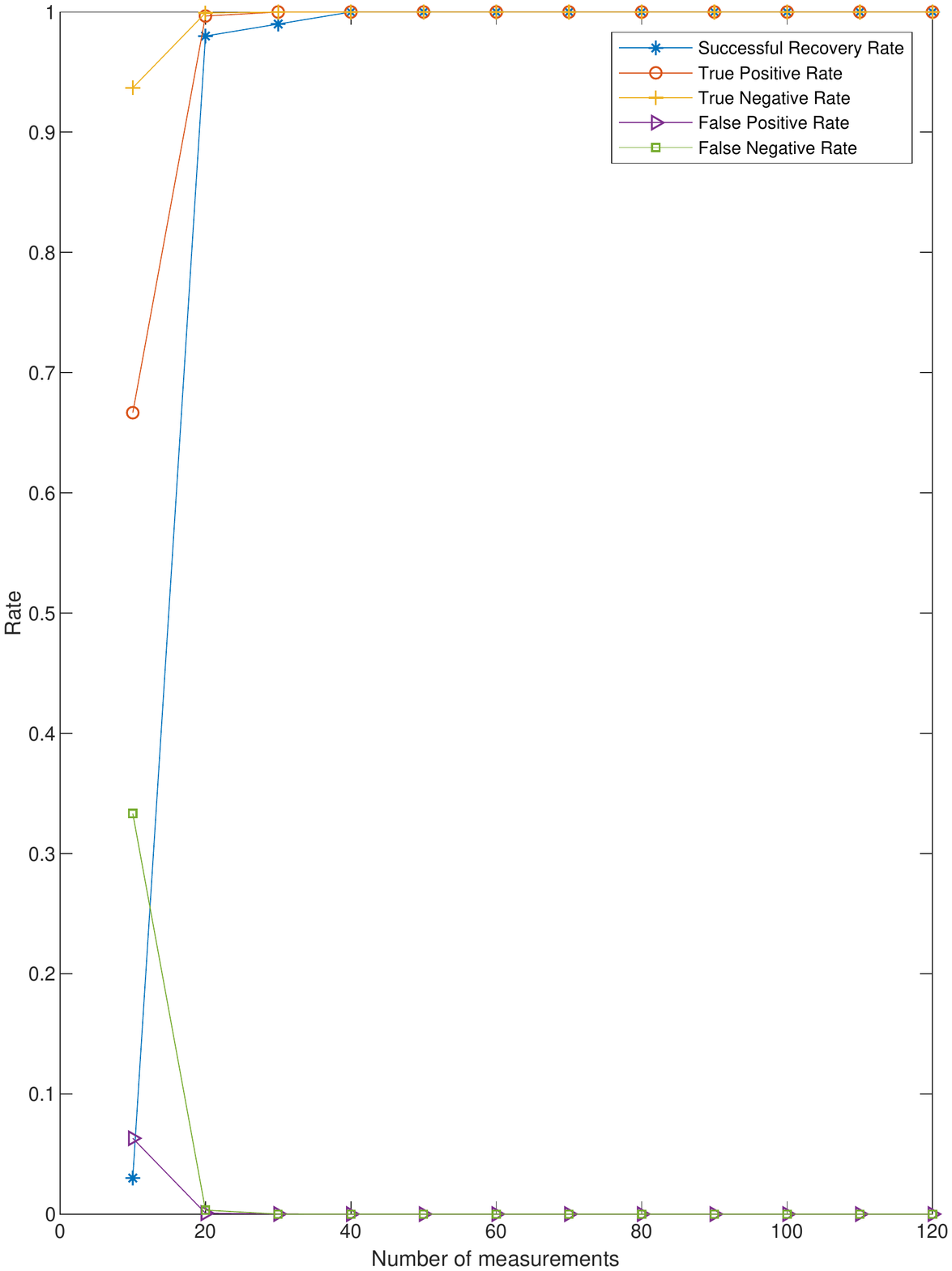}
		\caption{$n=120, k=3$.Expander measurement matrix with 5 '1's in each column. Noisy measurements.}
		\label{Fig:NoisyExpander3}
	\end{figure}

\begin{figure}
		\centering
		\includegraphics[width=0.5\textwidth, height=0.3\textwidth]{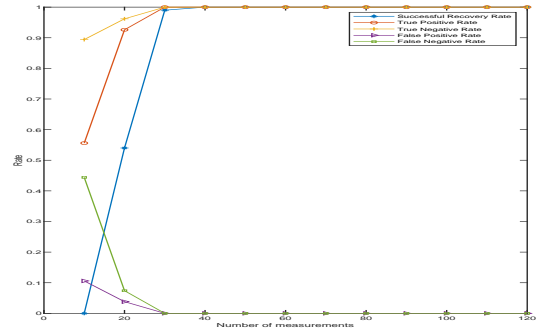}
		\caption{$n=120,k=5$.Expander measurement matrix with 5 '1's in each column. Noisy measurements.}
		\label{Fig:NoisyExpander4}
	\end{figure}

We also conduct experiments with noisy measurements, and the signal is recovered from noisy measurements by solving
\begin{align}
\min_{x\in\mbR^n} \|x\|_1, ~~{\rm s.t\ }\|Ax-y\|_2\leq \epsilon, x\geq 0,
\end{align}
where $\epsilon>0$ is a parameter tuned to noise magnitude, and $y\in\mbR^m$ is the noisy measurement vector. We follow the same setup as in previous section expect that for each trial of each set of parameter $(m,n)$, we add randomly generated noise vector $v$ with normalized magnitude $10^{-3}$ to the measurements, namely $y=Ax+v$. For each trial of each set of parameter, we treat the recovery as successful if it achieves a reconstruction error less than $10^{-2}$. The results of the recovery probabilities, false positive rates, and false negative rates are shown in the following figures from Figure \ref{Fig:NoisyBernoulli1} to Figure \ref{Fig:NoisyExpander4}. Figure \ref{Fig:ExpanderNoisy10fold} shows the results for $k=2$ and $n=200$, demonstrating a possible increase of throughput by 10 times.

We can see that similar increases in testing throughput are also observed as in the noiseless cases. In fact, for a large range of reasonable noise levels, we can observe similar increases in testing throughput with low false positive rates and false negative rates. 

\begin{figure}
		\centering
		\includegraphics[width=0.5\textwidth, height=0.3\textwidth]{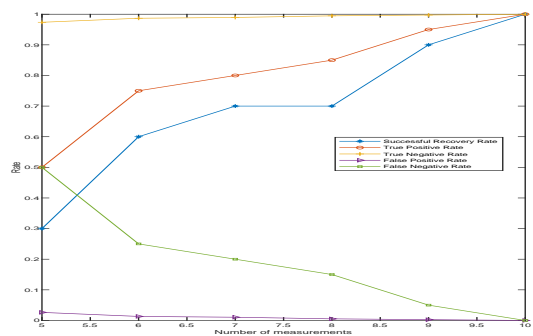}
	~ 

	\caption{Exhaustive search for binary measurement matrix with entries from Bernoulli distribution. The magnitude of the noise vector is set at $10^{-3}$. }\label{Fig:ExhaustSearch}
\end{figure}

In another experiment, we numerically evaluate the performance of exhaustive search in detecting viruses. We take $n=40$ and $k=2$, and the number of measurements is taken as  $m=5$,$6$,$7$,$8$, $9$, and $10$. For each set of $(m,n,k)$, we run 10 trials. In each trial, the pooling matrix is a Bernoulli random matrix. The measurement result is contaminated with random noise normalized to have a magnitude of $10^{-3}$.  A trial is considered to have successful recovery if the recovery error is less than $10^{-2}$ in the noisy case. In exhaustive search, since the true signal has sparsity of $k$, we will simply perform brute force calculations over all the possible sets of $k$ infected persons. For each possible such set of cardinality $k$, we  extract the corresponding columns from the measurement matrix. By doing this, we get an overdetermined system, and solve it via the least squares method. There are totally ${n \choose k}$ possible such sets, which means we need to solve the least square ${n \choose k}$ times for each trial. The results are shown in Figure \ref{Fig:ExhaustSearch}. As we can see, using only $10$ measurements, 
the false positive rate and false negative rates are very low (in fact 0 in this experiment). That amounts to a factor of $\frac{40}{10}=4$ speedup in throughput of the test. 

\begin{figure}
	\centering
		\includegraphics[width=0.5\textwidth, height=0.3\textwidth]{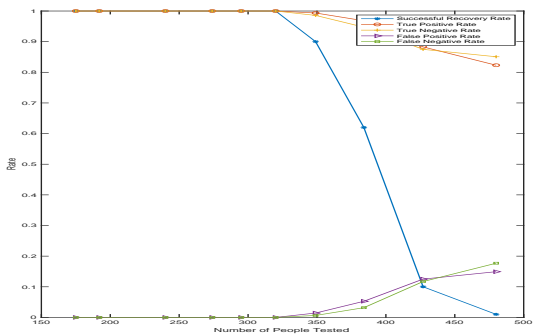}
		\caption{Rates versus Number of People Tested $n$. The number of pooling measurement is $m=96$, and $k \approx 0.087 \times n$ persons carry viruses. Binary measurement matrix with entries i.i.d. according to Bernoulli distribution.}
		\label{Fig:iowacase}
	\end{figure}

We now look at the testing data of COVID-19 viruses from the state of Iowa. The rate of testing positive is around 8.7 percent by early April, meaning among all the tests carried out, 8.7 percent of them came back with a `positive' result.  We consider a microplate of 96 wells,  and assume that the PCR machine can analyze 96 samples in one operational period.  Then we do a computational experiment to answer, ``using compressed sensing, for how many people these 96 compressed sensing (pooling) samples can correctly identify all the carriers of viruses present in that group of people?''   In this experiment, we fix the number of measurements, namely $m$,  as 96. Then we vary the number of people $n$, and randomly pick 8.7 percent of them (namely $k =ceil( 0.087 n)$, where $ceil(\cdot)$ is the ceiling function) as virus carriers.   We accordingly generate the virus quantity vector $x$. We plot the successful recovery rate of $x$, the false positive rate and false negative rate as functions of $n$ in Figure \ref{Fig:iowacase}. 
As $n$ increases, there are more virus carriers, and false positive rates and false negative rates are expected to increase when $m=96$ is fixed.  We observe that for $n \leq 300$, these false positive rates and false negative rates stay very low. This means that, when 8.7 percent of people have viruses, using compressed sensing,  the throughput of testing can grow to as much as $\frac{300}{96} \approx 3$ times.  For both Bernoulli random matrices and expander graph based matrices with 7 `1's in one column, we observe similar behaviors. 
When the percent of people carrying viruses decreases, say to $1$ percent, compressed sensing can even increase the throughput by more than 10 times. 

\begin{figure}
	\centering
		\includegraphics[width=0.5\textwidth, height=0.3\textwidth]{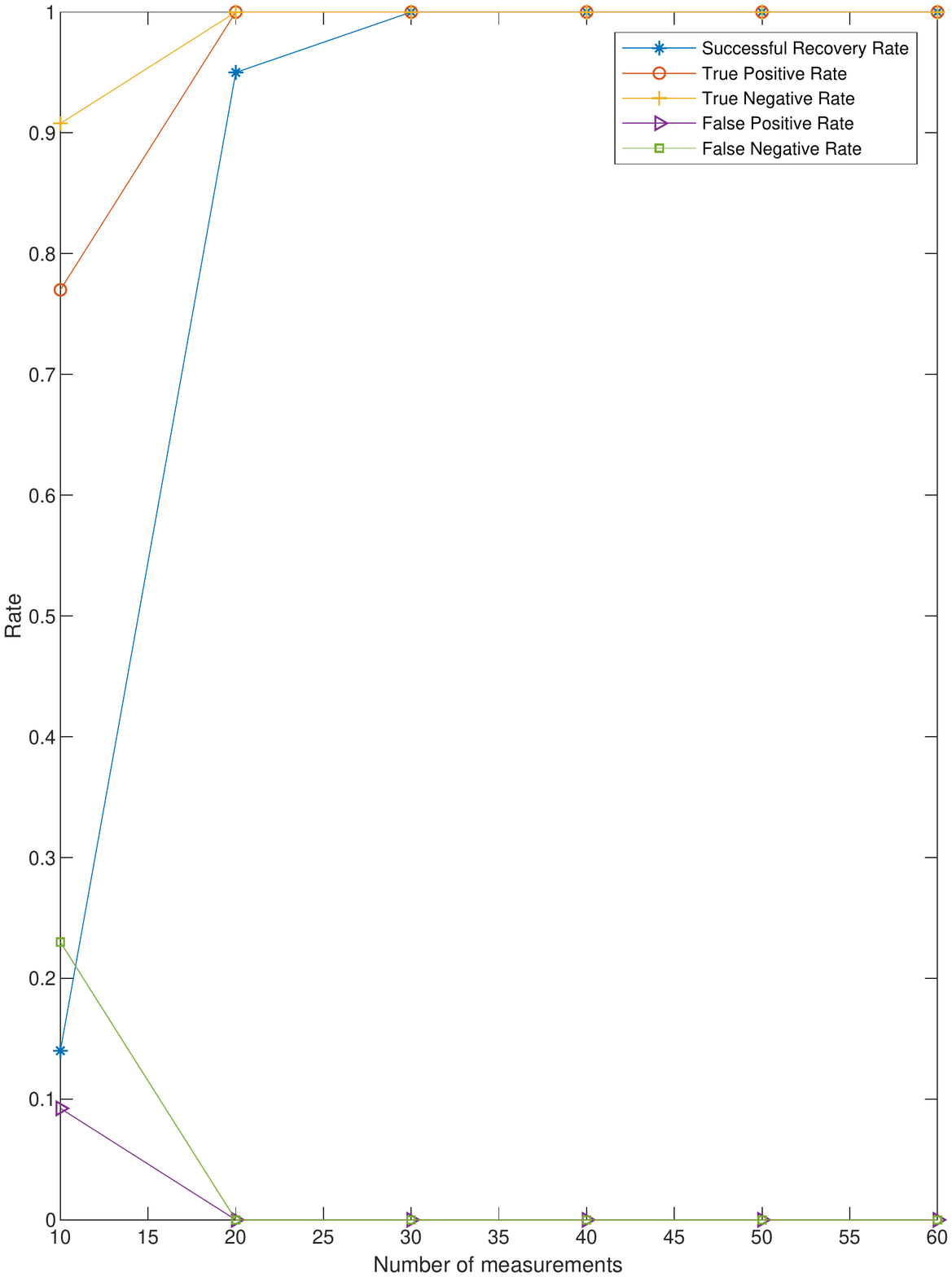}
		\caption{$n=60, k=3$. Binary measurement matrix with entries i.i.d. according to Bernoulli distribution. Noisy measurements.}
		\label{Fig:NoisyBernoulli1}
	\end{figure}%
	~ 
	
	~ 

	~ 

	~ 
\section{Discussions}
 This paper focuses on non-adaptive compressed sensing, which can have the advantange of minimizing the latency in obtaining the test results for tested persons. However, it is totally possible to increase the throughout of testing by using adpative measurements for compressed sensing, as adopted in \cite{poolingStanfordcommunity,Shani-Narkiss2020} for group testing.

\bibliographystyle{spmpsci}      
\bibliography{main}

\normalsize

\end{document}